\documentclass[longauth, traditabstract]{aa} 
\usepackage{graphicx}
\usepackage{txfonts}
\usepackage{natbib}
\begin{document}

\title{Sensitive limits on the abundance of cold water vapor in the DM
  Tau protoplanetary disk \thanks{Herschel is an ESA space observatory
    with science instruments provided by European-led Principal
    Investigator consortia and with participation important from
    NASA.}}

\author{
E.A.~Bergin\inst{\ref{inst1}}
\and M.R.~Hogerheijde\inst{\ref{inst2}}
\and C.~Brinch\inst{\ref{inst2}}
\and J.~Fogel\inst{\ref{inst1}}
\and U.A.~Y{\i}ld{\i}z\inst{\ref{inst2}}
\and L.E.~Kristensen\inst{\ref{inst2}}
%
\and E.F.~van~Dishoeck\inst{\ref{inst2},\ref{inst3}}
\and T.A.~Bell\inst{\ref{inst8}}
\and G.A.~Blake\inst{\ref{inst4}}
\and J.~Cernicharo\inst{\ref{inst5}}
\and C.~Dominik\inst{\ref{inst6},\ref{inst7}}
\and D.~Lis\inst{\ref{inst8}}
\and G.~Melnick\inst{\ref{inst9}}
\and D.~Neufeld\inst{\ref{inst10}}
\and O.~Pani\'c\inst{\ref{inst11}}
\and J.C.~Pearson\inst{\ref{inst12}}
%
\and R.~Bachiller\inst{\ref{inst13}}
\and A.~Baudry\inst{\ref{inst14}}
\and M.~Benedettini\inst{\ref{inst15}}
\and A.O.~Benz\inst{\ref{inst16}}
\and P.~Bjerkeli\inst{\ref{inst17}}
\and S.~Bontemps\inst{\ref{inst18}}
\and J.~Braine\inst{\ref{inst19}}
\and S.~Bruderer\inst{\ref{inst16}}
\and P.~Caselli\inst{\ref{inst19},\ref{inst15}}
\and C.~Codella\inst{\ref{inst15}}
\and F.~Daniel\inst{\ref{inst5}}
\and A.M.~di~Giorgio\inst{\ref{inst14}}
\and S.D.~Doty\inst{\ref{inst20}}
\and P.~Encrenaz\inst{\ref{inst21}}
\and M.~Fich\inst{\ref{inst22}}
\and A.~Fuente\inst{\ref{inst23}}
\and T.~Giannini\inst{\ref{inst24}}
\and J.R.~Goicoechea\inst{\ref{inst5}}
\and Th.~de~Graauw\inst{\ref{inst25}}
\and F.~Helmich\inst{\ref{inst25}}
\and G.J.~Herczeg\inst{\ref{inst3}}
\and F.~Herpin\inst{\ref{inst14}}
\and T.~Jacq\inst{\ref{inst15}}
\and D.~Johnstone\inst{\ref{inst26},\ref{inst27}}
\and J.K.~J{\o}rgensen\inst{\ref{inst28}}
\and B.~Larsson\inst{\ref{inst29}}
\and R.~Liseau\inst{\ref{inst17}}
\and M.~Marseille\inst{\ref{inst25}}
\and C.~M$^{\textrm c}$Coey\inst{\ref{inst22},\ref{inst30}}
\and B.~Nisini\inst{\ref{inst14}}
\and M.~Olberg\inst{\ref{inst17}}
\and B.~Parise\inst{\ref{inst31}}
\and R.~Plume\inst{\ref{inst32}}
\and C.~Risacher\inst{\ref{inst25}}
\and J.~Santiago-Garc\'{i}a\inst{\ref{inst33}}
\and P.~Saraceno\inst{\ref{inst14}}
\and R.~Shipman\inst{\ref{inst25}}
\and M.~Tafalla\inst{\ref{inst13}}
\and T.A.~van~Kempen\inst{\ref{inst9}}
\and R.~Visser\inst{\ref{inst2}}
\and S.F.~Wampfler\inst{\ref{inst16}}
\and F.~Wyrowski\inst{\ref{inst31}}
\and F.~van der Tak\inst{\ref{inst25},\ref{inst34}}
%
\and W.~Jellema\inst{\ref{inst25}}
\and A.G.G.M.~Tielens\inst{\ref{inst2}}
\and P.~Hartogh\inst{\ref{inst35}}
\and J.~St\"utzki\inst{\ref{inst36}}
\and R.~Szczerba\inst{\ref{inst37}}
}


\institute{
Department of Astronomy, The University of Michigan, 500 Church Street, Ann Arbor, MI 48109-1042, USA\label{inst1}
\and
Leiden Observatory, Leiden University, PO Box 9513, 2300 RA Leiden, The Netherlands\label{inst2}
\and
Max Planck Institut f\"{u}r Extraterrestrische Physik, Giessenbachstrasse 1, 85748 Garching, Germany\label{inst3}
\and
California Institute of Technology, Division of Geological and Planetary Sciences, MS 150-21, Pasadena, CA 91125, USA\label{inst4}
\and
Centro de Astrobiolog\'{\i}a. Departamento de Astrof\'{\i}sica. CSIC-INTA. Carretera de Ajalvir, Km 4, Torrej\'{o}n de Ardoz. 28850, Madrid, Spain.\label{inst5}
\and
Astronomical Institute Anton Pannekoek, University of Amsterdam, Kruislaan 403, 1098 SJ Amsterdam, The Netherlands\label{inst6}
\and
Department of Astrophysics/IMAPP, Radboud University Nijmegen, P.O. Box 9010, 6500 GL Nijmegen, The Netherlands\label{inst7}
\and
California Institute of Technology, Cahill Center for Astronomy and Astrophysics, MS 301-17, Pasadena, CA 91125, USA\label{inst8}
\and
Harvard-Smithsonian Center for Astrophysics, 60 Garden Street, MS 42, Cambridge, MA 02138, USA\label{inst9}
\and
Department of Physics and Astronomy, Johns Hopkins University, 3400 North Charles Street, Baltimore, MD 21218, USA\label{inst10}
\and
European Southern Observatory, Karl-Schwarzschild-Str. 2, 85748 Garching, Germany\label{inst11}
\and
Jet Propulsion Laboratory, California Institute of Technology, Pasadena, CA 91109, USA\label{inst12}
\and
Observatorio Astron\'{o}mico Nacional (IGN), Calle Alfonso XII,3. 28014, Madrid, Spain\label{inst13}
\and
INAF - Istituto di Fisica dello Spazio Interplanetario, Area di Ricerca di Tor Vergata, via Fosso del Cavaliere 100, 00133 Roma, Italy\label{inst14}
\and
INAF - Osservatorio Astrofisico di Arcetri, Largo E. Fermi 5, 50125 Firenze, Italy\label{inst15}
\and
Institute of Astronomy, ETH Zurich, 8093 Zurich, Switzerland\label{inst16}
\and
Department of Radio and Space Science, Chalmers University of Technology, Onsala Space Observatory, 439 92 Onsala, Sweden\label{inst17}
\and
Universit\'{e} de Bordeaux, Laboratoire d'Astrophysique de Bordeaux, France; CNRS/INSU, UMR 5804, Floirac, France\label{inst18}
\and
School of Physics and Astronomy, University of Leeds, Leeds LS2 9JT, UK\label{inst19}
\and
Department of Physics and Astronomy, Denison University, Granville, OH, 43023, USA\label{inst20}
\and
LERMA and UMR 8112 du CNRS, Observatoire de Paris, 61 Av. de l'Observatoire, 75014 Paris, France\label{inst21}
\and
University of Waterloo, Department of Physics and Astronomy, Waterloo, Ontario, Canada\label{inst22}
\and
Observatorio Astron\'{o}mico Nacional, Apartado 112, 28803 Alcal\'{a} de Henares, Spain\label{inst23}
\and
INAF - Osservatorio Astronomico di Roma, 00040 Monte Porzio catone, Italy\label{inst24}
\and
SRON Netherlands Institute for Space Research, PO Box 800, 9700 AV, Groningen, The Netherlands\label{inst25}
\and
National Research Council Canada, Herzberg Institute of Astrophysics, 5071 West Saanich Road, Victoria, BC V9E 2E7, Canada\label{inst26}
\and
Department of Physics and Astronomy, University of Victoria, Victoria, BC V8P 1A1, Canada\label{inst27}
\and
Centre for Star and Planet Formation, Natural History Museum of Denmark, University of Copenhagen,
{\O}ster Voldgade 5-7, DK-1350 Copenhagen K., Denmark\label{inst28}
\and
Department of Astronomy, Stockholm University, AlbaNova, 106 91 Stockholm, Sweden\label{inst29}
\and
The University of Western Ontario, Department of Physics and Astronomy, London, Ontario, N6A 3K7, Canada\label{inst30}
\and
Max-Planck-Institut f\"{u}r Radioastronomie, Auf dem H\"{u}gel 69, 53121 Bonn, Germany\label{inst31}
\and
Department of Physics and Astronomy, University of Calgary, Calgary, T2N 1N4, AB, Canada\label{inst32}
\and
Instituto de Radioastronom\'{i}a Milim\'{e}trica (IRAM), Avenida Divina Pastora 7, N\'{u}cleo Central, E-18012 Granada, Spain\label{inst33}
\and
Kapteyn Astronomical Institute, University of Groningen, PO Box 800, 9700 AV, Groningen, The Netherlands\label{inst34}
\and
Max-Planck-Insitut f\"ur Sonnensystemforschung, D 37191, Katlenburg-Lindau, Germany\label{inst35}
\and
KOSMA, I. Physik. Institut, Universit\"{a}t zu K\"{o}ln, Z\"{u}lpicher Str. 77, D 50937 K\"{o}ln, Germany\label{inst36}
\and
N. Copernicus Astronomical Center, Rabianska 8, 87-100, Torun, Poland\label{inst37}
}

\abstract{We performed a sensitive search for the ground-state
  emission lines of ortho- and para-water vapor in the DM Tau
  protoplanetary disk using the {\em Herschel/HIFI} instrument. No
  strong lines are detected down to 3$\sigma$ levels in
  0.5~km~s$^{-1}$ channels of 4.2~mK for the $1_{10}$--$1_{01}$ line
  and 12.6~mK for the $1_{11}$--$0_{00}$ line. We report a very
  tentative detection, however, of the $1_{10}$--$1_{01}$ line in the
  Wide Band Spectrometer, with a strength of $T_{\rm mb}=2.7$~mK, a
  width of 5.6 km~s$^{-1}$ and an integrated intensity of 16.0
  mK~km~s$^{-1}$. The latter constitutes a $6\sigma$
  detection. Regardless of the reality of this tentative detection,
  model calculations indicate that our sensitive limits on the line
  strengths preclude efficient desorption of water in the UV
  illuminated regions of the disk. We hypothesize that more than
  95--99\% of the water ice is locked up in coagulated grains that
  have settled to the midplane.}

   \keywords{ISM: abundances --- ISM: molecules --- protoplanetary disks}
   \titlerunning{Water Vapor in DM Tau}
	\authorrunning{Bergin et al.}
   \maketitle
%

\section{Introduction}

Because of its association with biology on Earth, water is one of the
most important molecules in the solar system and beyond.  However, the
origin of water on Earth is highly uncertain. 
What is clear is that the distribution of hydrated rocks in the solar
system \citep{abe_earthwater} suggests that water resided in the vapor
phase in the warm ($T_{dust} >$ 100 K) inner solar nebula
and is predominantly condensed in the form of ice beyond the so-called
snow-line between 2--2.5 AU \citep{hayashi_mmsn, abe_earthwater}.

The observation of cool water vapor in protoplanetary disks is
hampered by atmospheric attenuation and requires space-based
observations.  A number of recent detections of hot ($T_{\rm gas} >$
200~K) water by the {\em Spitzer} Space Telescope has demonstrated
that abundant water vapor is a ubiquitous component in
protoplanetary disks \citep{salyk08, cn08, pont10}.  However, this
emission likely arises within the snow-line of these young systems and
therefore does not provide a complete picture of the distribution of
water (both ice and vapor) in disks with sizes in excess of 100 AU.
In particular, these observations do not probe the cold ($T_{\rm
  dust}$ $\sim$ $T_{\rm gas} \sim 20$ K) outer parts of the disk where
most of the mass resides.

The {\em Herschel} Space Observatory \citep{pilbratt10}
offers a new
opportunity for a characterization of the distribution and evolution
of water vapor in protoplanetary disks.  We report here a sensitive
search for cold water emission in the ground-state lines of {\em
  ortho- (o)} and {\em para- (p)} H$_2$O using the high spectral
resolution of the \emph{HIFI} instrument \citep{degraauw10} towards
the well studied DM Tau protoplanetary disk. This study is part of the
guaranteed time key program `Water in Star Forming Regions' (van
Dishoeck et al. in prep.). Stringent limits to the strength of
both lines of a few mK suggest that the outer regions of the disk
contain little water vapor or water ice. In Sect.~2 we outline the
observations.  Section~3 presents results from detailed modeling;
Sect.~4 summarizes the implications of our result.

\section{Observations and results\label{s:obsres}}

DM Tau is a T Tauri star located at $\alpha(2000) =$
4$^h$33$^m$48\fs7 and $\delta(2000) =$ 18$^{\circ}$10$'$10$''$  
with a disk diameter, estimated from CO emission, of $\sim 1800$ AU at a systemic velocity of 6 km
s$^{-1}$ and $i \sim -35^{\circ}$ \citep{pdg07}.
The
source is a single M1 star \citep{wg01} with $L$=0.25~L$_{\odot}$.  DM Tau has a chemically rich molecular disk
\citep{dutrey_diskchem}.    Accretion from the
disk to the star also provides a source of excess UV
luminosity, 
with an overall UV field strength of $G_0 \sim 240$ \citep[relative to
the standard interstellar radiation field, ISRF;][]{bergin_lyalp,
  bergin_h2, habing68}.  The object is a transition disk
with an inner hole on the order of a few AU, based on models of {\em Spitzer} spectra  \citep{calvet_transdisks}.

DM~Tau was observed with the \emph{HIFI} instrument 
using the double beam switch observing mode with a throw of
$3{\farcm}0$. On 2010 March 22 spectra were taken in receiver band 1b
with an
on-source integration time of 198~minutes, and $T_{\rm sys}$=78--96 K.
On 2010 March 4 spectra were taken in receiver
band 4b
and an on-source integration time of 328~minutes, and $T_{\rm
  sys}$=370--410 K.

The \emph{HIFI} beam of $39''$ at 556~GHz and $21''$ at 1113~GHz is
larger than the DM~Tau disk with a diameter of
$12{\farcs}7$ at 140~pc. The beams are also larger than the pointing
accuracy of \emph{Herschel} of $\sim 2''$.
The data were recorded with Wide-Band Spectrometer
(WBS)
covering 4.4 GHz 
with 1.1 MHz resolution
(0.59 and 0.30 km~s$^{-1}$ at 556 and 1113 GHz, respectively), and the
High-Resolution Spectrometer (HRS)
covering 230 MHz 
at 0.25 MHz resolution (0.13 and 0.067 km~s$^{-1}$ at 556 and 1113
GHz, respectively). Both H- and V-polarizations were measured.

The raw data were calibrated onto the $T_A^*$ scale by the in-orbit
system and converted to $T_{\rm mb}$ assuming a
beam efficiency of 0.74 
The data were reduced using HIPE
v3.0. Subsequently, the data were exported to CLASS\footnote{{\tt
    http://www.iram.fr/IRAMFR/GILDAS}}. The \emph{HIFI} flux
calibration is accurate to 10\%, while the velocity scale of HIPE v3.0
is accurate to several m~s$^{-1}$. For both lines, the WBS data were
rebinned to 0.54 km~s$^{-1}$ channels (or 1.0 MHz, close to the
instrumental resolution of 1.1 MHz); the HRS data were rebinned to
0.45 km~s$^{-1}$ channels. 
All spectra, including the the H- and
V-polarizations, were averaged together weighted by their respective
noise levels.  The resulting rms noise levels are 2.9~mK (HRS) and
1.4~mK (WBS) for the H$_2$O $1_{01}$--$1_{01}$ line, and 7.2~mK (HRS)
and 4.3~mK (WBS) for the H$_2$O $1_{11}$--$0_{00}$
line\footnote{Although the channel spacing in both bands in similar
  (0.54 vs 0.45 km~s$^{-1}$), the noise in the WBS is 1.7--2.0 times
  lower than the noise in the HRS because of the larger noise
  bandwidth of the WBS and a $\sqrt{2}$ loss factor in the HRS
  autocorrelator.}. Figure~1 illustrates for band 1b that the noise in
our data decreases as (time)$^{-0.5}$ up to the full achieved
integration times.

Figure~2 presents the HRS and WBS spectra of the two water
transitions. No strong lines are detected; for comparison, Fig.~2 also
shows the 
$^{12}$CO 1--0 spectrum of DM~Tau (\citealt{kesslerphd};
Pani\'c et al. in prep.) showing a clear emission line with a
width of $\sim$2 km~s$^{-1}$ 
%
%
%
centered on the source velocity of $+6.1$ km~s$^{-1}$. In the WBS
spectrum of the H$_2$O $1_{10}$--$1_{01}$ line a weak feature is
present between $V_{\rm LSR}$ +0.5 and +10 km~s$^{-1}$, peaking around
+6.6 km~s$^{-1}$; a similar feature is seen in the noisier HRS
spectrum. With $T_{\rm mb}=4.3$~mK, the brightest channel lies at
3$\sigma$. Integrated between +0.5 and +10 km~s$^{-1}$, the feature
contains $17.2\pm 3.2$ mK~km~s$^{-1}$, a 5$\sigma$ result. A Gaussian
fit to the feature yields best fit parameters of $V_{\rm LSR}=+6.8\pm
0.7$ km~s$^{-1}$, a FWHM width of $5.6\pm 1.2$ km~s$^{-1}$, an
intensity $T_{\rm mb}=2.7\pm 1.4$~mK, and an integrated intensity of
$16.0\pm 2.7$ mK~km~s$^{-1}$ (6$\sigma$).

Arguments in favor of interpreting this feature as a positive
detection of the H$_2$O $1_{10}$--$1_{01}$ line include the facts that
the integrated intensity constitutes a 5--6$\sigma$ detection and that
the feature peaks near the systemic velocity of 6.1 km~s$^{-1}$.
Against the interpretation as a positive detection is
a line peak that is only 2--3$\sigma$ and a linewidth which is twice that of the $^{12}$CO line. The width would suggest that the
emission arises from within 10~AU. With respect to the latter, it is
interesting that HCO$^+$ 1--0 line 
has a blue wing extending over $\sim$3 km~s$^{-1}$ \citep{dutrey_diskchem}.

Neither set of arguments is clearly stronger, and we interpret the
feature in the H$_2$O $1_{10}$--$1_{01}$ spectrum as a \emph{very
  tentative} detection of water vapor in the disk of DM~Tau.
In the remainder of this Letter, we will work with an intensity of
$\lesssim 2.7$~mK for the H$_2$O $1_{10}$--$1_{01}$ line and with an
upper limit to the intensity of the H$_2$O $1_{11}$--$0_{00}$ line of
$<12.6$~mK.

Smoothing the WBS spectra to $\sim 27$ km~s$^{-1}$ resolution results
in positive detections of the continuum of DM~Tau at 556.9~GHz of $3.0\pm 0.5$ mK ($0.6\pm 0.1$ Jy) and at 1113.3~GHz of $5.7\pm 0.9$ mK ($1.3\pm 0.2$ Jy). These values are
consistent with other continuum measurements
\citep{dutrey96}.

\begin{figure}
\includegraphics[angle=0,width=0.95\columnwidth]{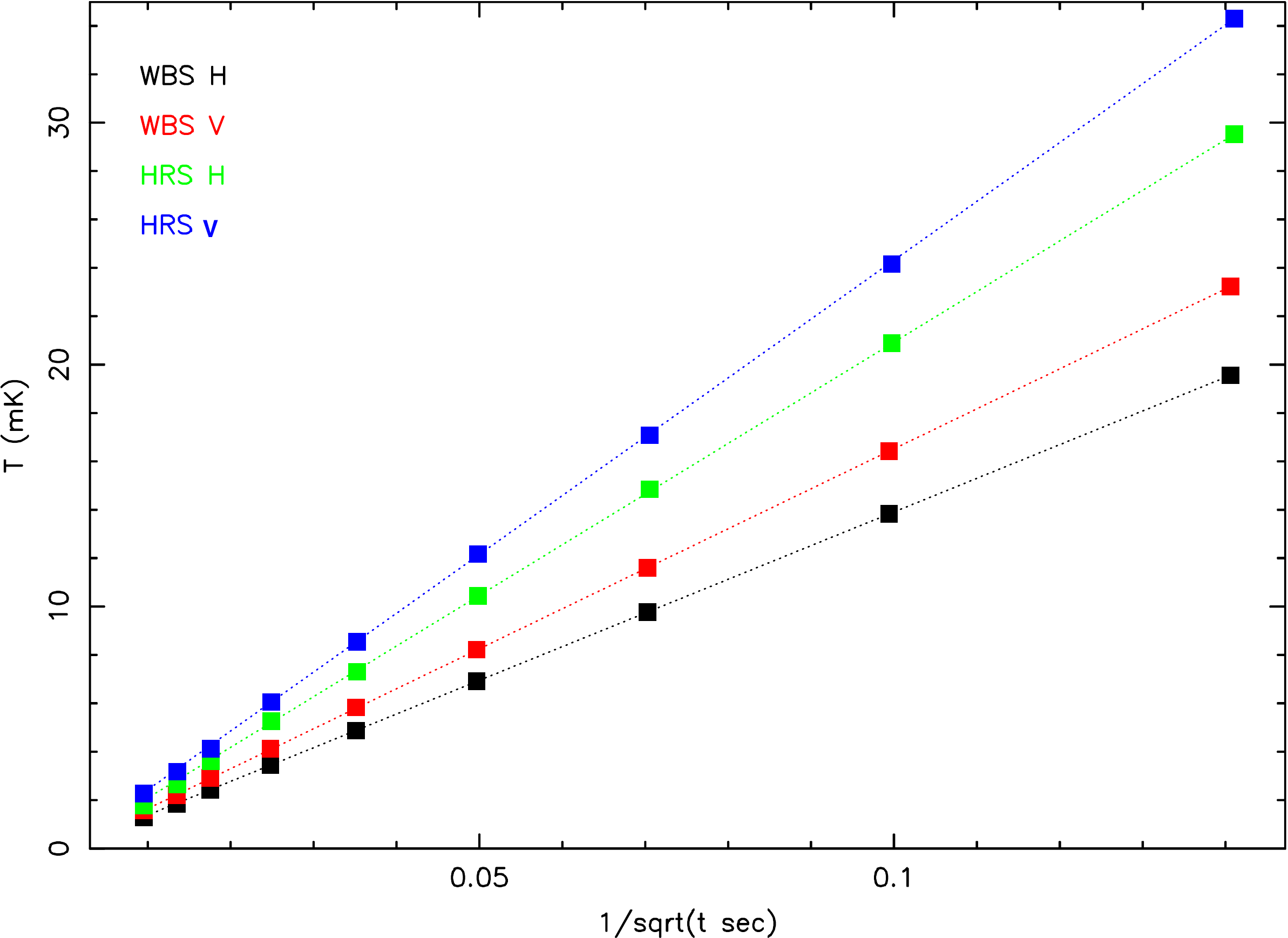}
\caption{Noise level (rms) vs integration time of the WBS and HRS data
  of the H$_2$O $1_{10}$-$1_{01}$ line. The noise decreases as
  $\sqrt{t_{\rm int}}$.}
\end{figure}

\begin{figure}
\includegraphics[width=0.95\columnwidth]{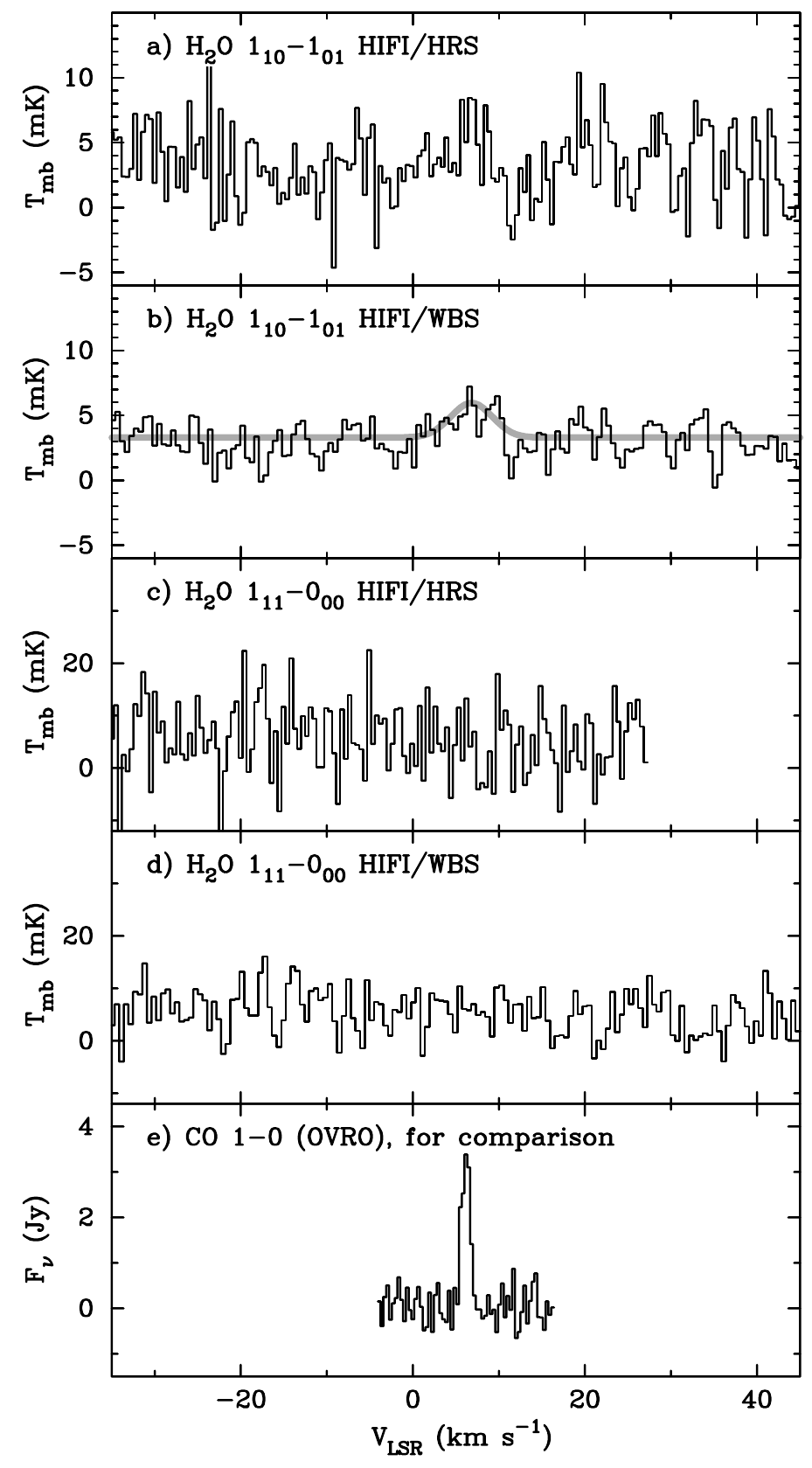}
\caption{(a)--(d) H$_2$O spectra obtained with \emph{HIFI}. A 
  tentative feature is present in panel (b) for H$_2$O
  $1_{10}$--$1_{01}$ detected with WBS between $V_{\rm LSR}$ of $+0.5$
  and +10 km~s$^{-1}$. A Gaussian fit to the feature is shown by the
  smooth grey line. The HRS data on this line in panel (a) are
  consistent with this result but noisier. Only noise is seen in
  panels (c) and (d) showing the $1_{11}$--$0_{00}$ line. (e)
  Comparison $^{12}$CO 1--0 spectrum of DM Tau obtained with OVRO
  (\citealt{kesslerphd}; Pani\'c et al. in prep.). 
}
\end{figure}

\section{Model Predictions}

\subsection{Chemistry of Water Vapor in the Cold Outer Disk}

Because the midplane temperatures are well below the evaporation
temperature at densities representative of the midplane of $>$150 K
\citep{fraser_h2obind, dcw05}, we can expect that beyond the snow-line
water is mostly frozen on the surfaces of dust grains.
Therefore, molecular emission arises predominantly from the warm disk
surface that is heated by stellar irradiation \citep{aikawa_vanz02}.
Beyond 10 AU the dust in this superheated layer is heated to $<$100 K
\citep{nomura07} and again it is anticipated that water will remain as
ice.  However, the upper layers of the disk are exposed to energetic
X-ray and FUV radiation which provide a source for non-thermal
desorption.  Models of X-ray induced desorption suggest that X-rays
cannot release significant H$_2$O into the gas \citep{najita_xray}.  A
more profitable method to desorb water ice in the cold outer disk is
via photodesorption, which has a measured yield (molecules/photon) of
$\sim 10^{-3}$ \citep{olvv10}.  This has been suggested as providing a
basal column of water vapor in the disk by \citet{dchk05}.

\begin{figure}
   \centering
   \includegraphics[width=0.95\columnwidth]{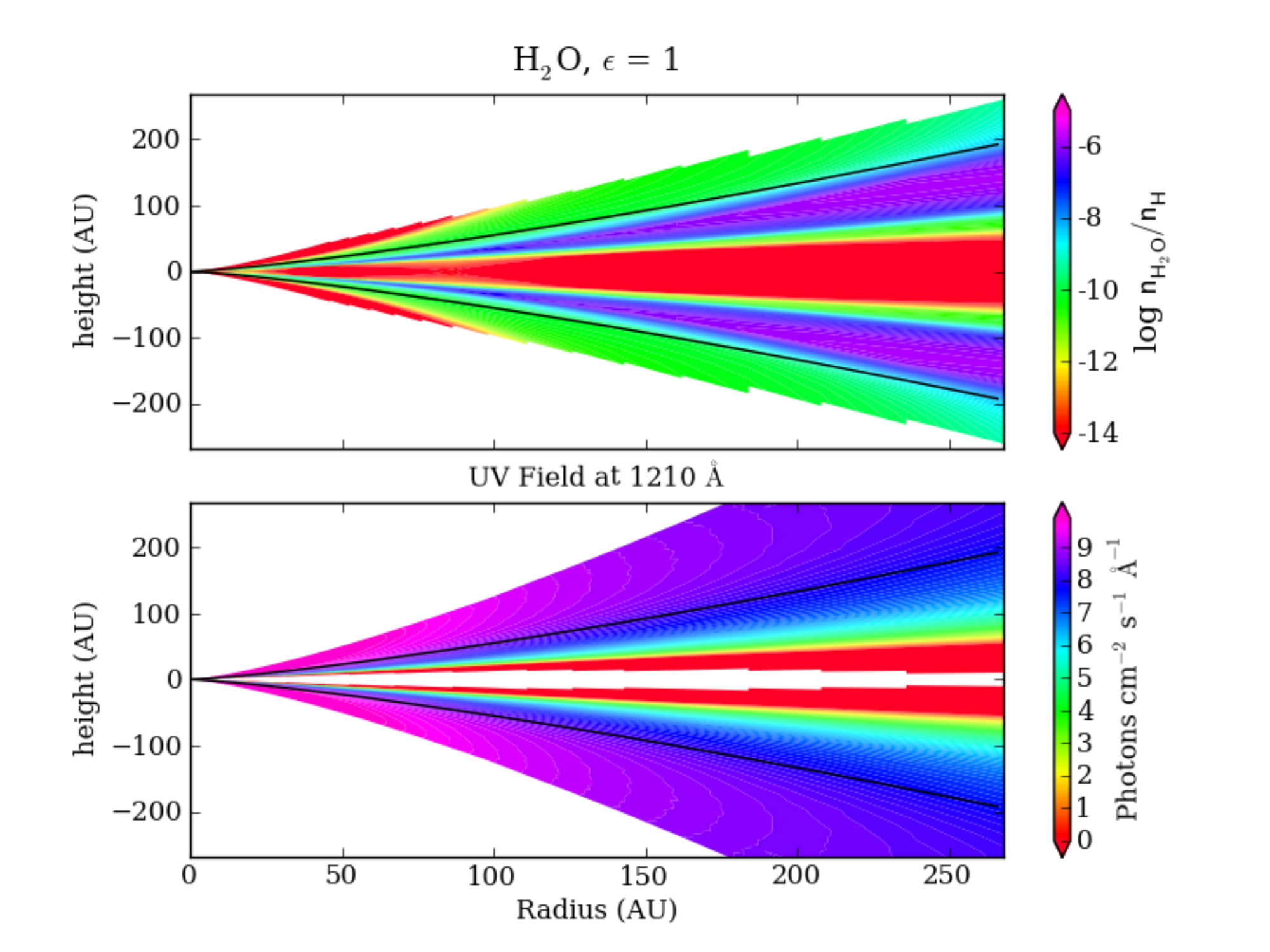}
   \caption{
   Abundance of water vapor ({\em Top})  and Ly $\alpha$ photon flux ({\em Bottom}) shown as a function of both radial distance and vertical height. The water vapor abundance is relative to total hydrogen and is only plotted for points where the vertical height to radial distance ratio is less than 1. The line in both plots refers to the $\tau$ = 1 surface for stellar radiation and $\epsilon$ = 1 refers to the fact that these models were run with a standard dust to gas ratio. All model results taken from \citet{fogel10}.  }
         \label{fig:chem}
\end{figure}

Figure~\ref{fig:chem} presents results from a detailed calculation of
the UV radiation transfer and chemistry of a standard T Tauri disk
with $M_{\rm disk}=0.03$ M$_\odot$, $R_{\rm out}=400$ AU and a
standard gas-to-dust ratio \citep{fogel10}.  A key aspect in the
calculation of the importance of photodesorption is the formation of
water ice on the grain surface via oxygen hydrogenation and also the
2D propagation of UV photons.  This includes Ly $\alpha$ radiation.
which is important for H$_2$O \citep{vd_diskphoto}.  As can be
seen a thin layer exists where water vapor is present in moderate
abundance ($\sim 10^{-7} - 10^{-6}$) in the cold outer regions of the
disk.


As noted by \citet{dchk05} the column density of water vapor produced
by photodesorption is independent of the photon flux.  This can be seen by balancing formation by photodesorption with destruction by UV photodissociation, giving a maximum
water abundance \citep{hkbm08}:
$n({\rm H_2O})\int \sigma_{\lambda}F_{\lambda}d\lambda = F_{\rm UV}Y\sigma_{\rm site}N_{\rm p}f_{\rm ice}N_{\rm m}^{-1}n({\rm H_2O})_{\rm ice}.
$
Here $F_{\rm UV} = \int F_{\lambda}d\lambda$ is the integral of the
photon flux given in Fig.~\ref{fig:chem} and $\sigma_{\lambda} \sim \sigma_{\rm
  Ly\;\alpha}\,\sim\,10^{-17}$ cm$^{-2}$
\citep{vd_diskphoto}. $\sigma_{\rm site}$ is the cross section of a
given site on the grain ($= \pi a^2/N_{\rm site}$, with grain radius $a$=0.1 $\mu$m
and $N_{\rm site}$=$10^6$).  $N_{\rm p}$=2 is a correction for the
fact the UV photons only penetrate the first few monolayers
\citep{olvv10}, with a yield of $Y \sim 2 \times 10^{-3}$.  $f_{\rm ice}$ is the fraction of water ice over the
total amount of ice and $N_{\rm m}$ is the number of monolayers.  Based on this approximation we find  $x({\rm H_2O})_{max} \sim\,10^{-6}$; for scaling relations see \citet{hkbm08}.  This could be lowered if the
grains have a reduced fraction of water ice or perhaps less than a
monolayer of coverage.

\subsection{Comparison to Observations}

Using the above chemical model calculation as input, we use the
molecular excitation and radiative transfer code LIME (Brinch et
al. in prep.) to calculate the line intensity in both observed water
lines.  We use the collision rates of water with {\it p}-H$_2$ from
\citet{faure2007} as provided by the LAMDA database
\citep{schoier2005}\footnote{\tt
  http://www.strw.leidenuniv.nl/$\sim$moldata} and convolved the
results with the appropriate \emph{Herschel} beams.  In our models we assume an intrinsic
broadening of 0.4 km/s on top of a Keplerian velocity profile.   

Because of the appreciable abundance of H$_2$O in the model
(disk-averaged column density of $6\times 10^{14}$
cm$^{-2}$), it is not surprising that significant line intensities are
predicted of $T_{\rm mb}=140$~mK for the $1_{10}$--$1_{01}$ line and
300~mK for the $1_{11}$--$0_{00}$ line  with a Gaussian spectral profile.   Clearly, our observations rule
out the presence of the amounts of water vapor predicted by
photodesorption regardless of the details of our model.

Absorption by low-excitation water from foreground material cannot
explain the absence of detected emission: the cloud seen in $^{12}$CO
by \citep{dutrey_diskchem} has narrow emission centered 3 km~s$^{-1}$ away
from the DM~Tau disk. Only when the column density of water is scaled
down by a factor of 130 to a disk-averaged value of $5\times 10^{12}$
cm$^{-2}$ does the predicted strength of the $1_{10}$--$1_{01}$ line
becomes consistent with the observed limits; the limits on the
$1_{11}$--$0_{00}$ line are less strict because of the higher noise of
these observations.

Our model could predict lines that are too intense if we overestimate
the collisional excitation of water. 
\citet{ddp10} suggested that existing collisional excitation rates for water are overestimated at temperatures below $\sim$50--80~K.   
Decreasing the collisional
excitation rates has little effect on the line strengths for the
original column density. In that case, the lines are still highly
optically thick, with maximum optical depth of 2000, and line trapping effectively excites the
line. However, for collision rates lower by a factor of 10 compared to
the adopted rates, reducing the disk averaged column density by a
factor of 20 to $3\times 10^{13}$ cm$^{-2}$ is sufficient to comply
with the observational constraints.  

Our generic model disk contains 0.03 M$_\odot$ similar to DM~Tau, but
its 400 AU radius is only half that of DM~Tau. The increased beam
dilution only strengthens our conclusions. For a DM~Tau specific
model, \citet{dchk05} predict the $1_{10}$--$1_{01}$ line to be in
absorption. Line profiles with combinations of emission and strong
absorption naturally arise in disk models with strong temperature,
density and water abundance gradients viewed under non-zero
inclinations. An example is provided by \citet{cernicharo2009} for the
HD 97048 disk. These models have water abundances peaking at much
larger height above the midplane, explaining the low excitation
resulting in strong absorption. For our discussion here, we stress
that all these models predict emission or absorption lines that are
inconsistent with our observational limits. Future work will focus on
more detailed models for DM~Tau, and also explore the effect of dust
settling and non-standard gas-to-dust ratios (Brinch et al. in prep.).

Finally, we note that \citet{cecc_hdodisk}
  have claimed a 4$\sigma$ detection HDO in the DM Tau.  This
  detection has been cast into doubt based on line formation
  considerations \citep{guilloteau_hdodisk}.  Our models predict water will be
  in emission, and not absorption.  At face value this would argue against the reality of the absorption.  If we derive a column density from the HDO observations and the limits here, the D/H ratio would also be exceedingly high $> 0.2$.
  Deeper HDO observations
are needed to settle this issue.

\section{Implications}

We have presented the results from a deep search for the ground state
emission lines of $o-$H$_2$O and $p-$H$_2$O towards the DM Tau disk.
Based on the best theoretical knowledge we have to date, water vapor
should be present in the outer disk and presumably emissive.  However,
our sensitive observations show that, at least for this object, it is
not. Our limit on the $p$-H$_2$O line precludes an extreme
$o/p$ ratio as explanation for the low $o$-H$_2$O emission strength.
There are two potential explanations for this result.  Either water
vapor is unemissive as would be the case if the excitation at low
temperature is lower than generally assumed
\citep{ddp10}. \citet{cernicharo2009} also show that the water
excitation sensitively depends on the adopted collision rates with
{\it o}- and {\it p}-H$_2$. However, our calculations suggest that
line trapping is sufficiently effective at the predicted water
abundances to still produce detectable lines. Alternatively, our
physical/chemical understanding may be incorrect.  The photodesorption
yield is measured in the laboratory at low temperature and the
abundance is independent of the photon flux so these aspects appear
unlikely to provide the answer.  One intriguing possibility is that
the upper layers of the outer disk are `dry' which could be the case
if only bare grains are present in the region where UV photons are
present.  A well known key aspect of disk physical evolution is the
coagulation and settling of dust grains to the disk midplane
\citep{wc_ppiii, dd04, furlan2006}.  Icy grains present a more
favorable surface for grain coagulation and would therefore become
larger and settle to deeper layers than their bare silicate
counterparts \citep{dominik1997}. While upward mixing of gas and small
grains may occur, larger ice-bearing grains remain in the midplane.
In addition, the total $n_{\rm grain}\,\sigma_{\rm grain}$ will be reduced, thereby
reducing the efficiency of grain surface formation of H$_2$O upon
which the photodesorption model depends for water vapor creation
\citep{hkbm08}.  This `cold-finger' effect was also proposed by
\citet{meijerink2009} to explain the truncation of warm
water vapor beyond $\sim$1~AU seen in \emph{Spitzer}
measurements. Thus, \emph{Herschel} and \emph{Spitzer} both suggest
that the disk around DM~Tau is settled. In summary, our
\emph{Herschel} results suggest that less than 1--5\% of the water ice
reservoir survives in the UV-illuminated \emph{outer} disk regions
around DM~Tau.  If this finding is confirmed by more detailed models
and by additional observations, \emph{Herschel} may be
telling us something entirely new about the chemical structure of
protoplanetary disks.

\begin{acknowledgements}
  \emph{HIFI} has been designed and built by a consortium of
  institutes and university departments from across Europe, Canada and
  the United States under the leadership of SRON Netherlands Institute
  for Space Research, Groningen, The Netherlands and with major
  contributions from Germany, France and the US.  Consortium members
  are: Canada: CSA, U.Waterloo; France: CESR, LAB, LERMA, IRAM;
  Germany: KOSMA, MPIfR, MPS; Ireland, NUI Maynooth; Italy: ASI,
  IFSI-INAF, Osservatorio Astrofisico di Arcetri- INAF; Netherlands:
  SRON, TUD; Poland: CAMK, CBK; Spain: Observatorio Astron{\'o}mico
  Nacional (IGN), Centro de Astrobiolog{\'\i}a (CSIC-INTA). Sweden:
  Chalmers University of Technology - MC2, RSS \& GARD; Onsala Space
  Observatory; Swedish National Space Board, Stockholm University -
  Stockholm Observatory; Switzerland: ETH Zurich, FHNW; USA: Caltech,
  JPL, NHSC.  Support for this work was provided by NASA through an
  award issued by JPL/Caltech.  EAB acknowledges support by NSF Grant
  0707777, MRH by NWO grant 639.042.404.

\end{acknowledgements}

\bibliography{ted}

\begin{thebibliography}{37}
\expandafter\ifx\csname natexlab\endcsname\relax\def\natexlab#1{#1}\fi

\bibitem[{{Abe} {et~al.}(2000){Abe}, {Ohtani}, {Okuchi}, {Righter}, \&
  {Drake}}]{abe_earthwater}
{Abe}, Y., {Ohtani}, E., {Okuchi}, T., {Righter}, K., \& {Drake}, M. 2000,
  {Water in the Early Earth} (Origin of the earth and moon, edited by
  R.M.~Canup and K.~Righter and 69 collaborating authors.~Tucson: University of
  Arizona Press., p.413-433), 413--433

\bibitem[{{Aikawa} {et~al.}(2002){Aikawa}, {van Zadelhoff}, {van Dishoeck}, \&
  {Herbst}}]{aikawa_vanz02}
{Aikawa}, Y., {van Zadelhoff}, G.~J., {van Dishoeck}, E.~F., \& {Herbst}, E.
  2002, \aap, 386, 622

\bibitem[{{Bergin} {et~al.}(2003){Bergin}, {Calvet}, {D'Alessio}, \&
  {Herczeg}}]{bergin_lyalp}
{Bergin}, E., {Calvet}, N., {D'Alessio}, P., \& {Herczeg}, G.~J. 2003, \apjl,
  591, L159

\bibitem[{{Bergin} {et~al.}(2004){Bergin}, {Calvet}, {Sitko}, {Abgrall},
  {D'Alessio}, {Herczeg}, {Roueff}, {Qi}, {Lynch}, {Russell}, {Brafford}, \&
  {Perry}}]{bergin_h2}
{Bergin}, E., {Calvet}, N., {Sitko}, M.~L., {et~al.} 2004, \apjl, 614, L133

\bibitem[{{Calvet} {et~al.}(2005)}]{calvet_transdisks}
{Calvet}, N. {et~al.} 2005, \apjl, 630, L185

\bibitem[{{Carr} \& {Najita}(2008)}]{cn08}
{Carr}, J.~S. \& {Najita}, J.~R. 2008, Science, 319, 1504

\bibitem[{{Ceccarelli} {et~al.}(2005){Ceccarelli}, {Dominik}, {Caux},
  {Lefloch}, \& {Caselli}}]{cecc_hdodisk}
{Ceccarelli}, C., {Dominik}, C., {Caux}, E., {Lefloch}, B., \& {Caselli}, P.
  2005, \apjl, 631, L81

\bibitem[{{Cernicharo} {et~al.}(2009){Cernicharo}, {Ceccarelli}, {M{\'e}nard},
  {Pinte}, \& {Fuente}}]{cernicharo2009}
{Cernicharo}, J., {Ceccarelli}, C., {M{\'e}nard}, F., {Pinte}, C., \& {Fuente},
  A. 2009, \apjl, 703, L123

\bibitem[{{D'Alessio} {et~al.}(2005){D'Alessio}, {Calvet}, \& {Woolum}}]{dcw05}
{D'Alessio}, P., {Calvet}, N., \& {Woolum}, D.~S. 2005, in Astronomical Society
  of the Pacific Conference Series, Vol. 341, Chondrites and the Protoplanetary
  Disk, ed. A.~N. {Krot}, E.~R.~D. {Scott}, \& B.~{Reipurth}, 353--+

\bibitem[{{de Graauw} {et~al.}(2010){de Graauw}, {Helmich}, {Phillips},
  {et~al.}}]{degraauw10}
{de Graauw}, T., {Helmich}, F.~P., {Phillips}, T., {et~al.} 2010, \aap, 518, L6

\bibitem[{{Dick} {et~al.}(2010){Dick}, {Drouin}, \& {Pearson}}]{ddp10}
{Dick}, M.~J., {Drouin}, B.~J., \& {Pearson}, J.~C. 2010, \pra, 81, 022706

\bibitem[{{Dominik} {et~al.}(2005){Dominik}, {Ceccarelli}, {Hollenbach}, \&
  {Kaufman}}]{dchk05}
{Dominik}, C., {Ceccarelli}, C., {Hollenbach}, D., \& {Kaufman}, M. 2005,
  \apjl, 635, L85

\bibitem[{{Dominik} \& {Tielens}(1997)}]{dominik1997}
{Dominik}, C. \& {Tielens}, A.~G.~G.~M. 1997, \apj, 480, 647

\bibitem[{{Dullemond} \& {Dominik}(2004)}]{dd04}
{Dullemond}, C.~P. \& {Dominik}, C. 2004, \aap, 421, 1075

\bibitem[{{Dutrey} {et~al.}(1996){Dutrey}, {Guilloteau}, {Duvert}, {Prato},
  {Simon}, {Schuster}, \& {Menard}}]{dutrey96}
{Dutrey}, A., {Guilloteau}, S., {Duvert}, G., {et~al.} 1996, \aap, 309, 493

\bibitem[{{Dutrey} {et~al.}(1997){Dutrey}, {Guilloteau}, \&
  {Guelin}}]{dutrey_diskchem}
{Dutrey}, A., {Guilloteau}, S., \& {Guelin}, M. 1997, \aap, 317, L55

\bibitem[{{Faure} {et~al.}(2007){Faure}, {Crimier}, {Ceccarelli}, {Valiron},
  {Wiesenfeld}, \& {Dubernet}}]{faure2007}
{Faure}, A., {Crimier}, N., {Ceccarelli}, C., {et~al.} 2007, \aap, 472, 1029

\bibitem[{{Fogel} {et~al.}(2010){Fogel}, {Bethell}, {Bergin}, {Calvet}, \&
  {Semenov}}]{fogel10}
{Fogel}, J.~K.~J., {Bethell}, T.~J., {Bergin}, E.~A., {Calvet}, N., \&
  {Semenov}, D. 2010, ApJ, submitted

\bibitem[{{Fraser} {et~al.}(2001){Fraser}, {Collings}, {McCoustra}, \&
  {Williams}}]{fraser_h2obind}
{Fraser}, H.~J., {Collings}, M.~P., {McCoustra}, M.~R.~S., \& {Williams}, D.~A.
  2001, \mnras, 327, 1165

\bibitem[{{Furlan} {et~al.}(2006){Furlan}, {Hartmann}, {Calvet}, {D'Alessio},
  {Franco-Hern{\'a}ndez}, {Forrest}, {Watson}, {Uchida}, {Sargent}, {Green},
  {Keller}, \& {Herter}}]{furlan2006}
{Furlan}, E., {Hartmann}, L., {Calvet}, N., {et~al.} 2006, \apjs, 165, 568

\bibitem[{{Guilloteau} {et~al.}(2006){Guilloteau}, {Pi{\'e}tu}, {Dutrey}, \&
  {Gu{\'e}lin}}]{guilloteau_hdodisk}
{Guilloteau}, S., {Pi{\'e}tu}, V., {Dutrey}, A., \& {Gu{\'e}lin}, M. 2006,
  \aap, 448, L5

\bibitem[{{Habing}(1968)}]{habing68}
{Habing}, H.~J. 1968, \bain, 19, 421

\bibitem[{{Hayashi}(1981)}]{hayashi_mmsn}
{Hayashi}, C. 1981, Progress of Theoretical Physics Supplement, 70, 35

\bibitem[{{Hollenbach} {et~al.}(2009){Hollenbach}, {Kaufman}, {Bergin}, \&
  {Melnick}}]{hkbm08}
{Hollenbach}, D., {Kaufman}, M.~J., {Bergin}, E.~A., \& {Melnick}, G.~J. 2009,
  \apj, 690, 1497

\bibitem[{{Kessler-Silacci}(2004)}]{kesslerphd}
{Kessler-Silacci}, J. 2004, PhD thesis, Caltech

\bibitem[{{Meijerink} {et~al.}(2009){Meijerink}, {Pontoppidan}, {Blake},
  {Poelman}, \& {Dullemond}}]{meijerink2009}
{Meijerink}, R., {Pontoppidan}, K.~M., {Blake}, G.~A., {Poelman}, D.~R., \&
  {Dullemond}, C.~P. 2009, \apj, 704, 1471

\bibitem[{{Najita} {et~al.}(2001){Najita}, {Bergin}, \& {Ullom}}]{najita_xray}
{Najita}, J., {Bergin}, E.~A., \& {Ullom}, J.~N. 2001, \apj, 561, 880

\bibitem[{{Nomura} {et~al.}(2007){Nomura}, {Aikawa}, {Tsujimoto}, {Nakagawa},
  \& {Millar}}]{nomura07}
{Nomura}, H., {Aikawa}, Y., {Tsujimoto}, M., {Nakagawa}, Y., \& {Millar}, T.~J.
  2007, \apj, 661, 334

\bibitem[{{{\"O}berg} {et~al.}(2009){{\"O}berg}, {Linnartz}, {Visser}, \& {van
  Dishoeck}}]{olvv10}
{{\"O}berg}, K.~I., {Linnartz}, H., {Visser}, R., \& {van Dishoeck}, E.~F.
  2009, \apj, 693, 1209

\bibitem[{{Pi{\'e}tu} {et~al.}(2007){Pi{\'e}tu}, {Dutrey}, \&
  {Guilloteau}}]{pdg07}
{Pi{\'e}tu}, V., {Dutrey}, A., \& {Guilloteau}, S. 2007, \aap, 467, 163

\bibitem[{{Pilbratt} {et~al.}(2010){Pilbratt}, {Riedinger}, {Passvogel},
  {et~al.}}]{pilbratt10}
{Pilbratt}, G., {Riedinger}, J.~R., {Passvogel}, T., {et~al.} 2010, \aap, 518,
  L1

\bibitem[{{Pontoppidan} {et~al.}(2010){Pontoppidan}, {Salyk}, {Blake},
  {Meijerink}, {Carr}, \& {Najita}}]{pont10}
{Pontoppidan}, K.~M., {Salyk}, C., {Blake}, G.~A., {et~al.} 2010, \apj, 00, in
  press

\bibitem[{{Salyk} {et~al.}(2008){Salyk}, {Pontoppidan}, {Blake}, {Lahuis}, {van
  Dishoeck}, \& {Evans}}]{salyk08}
{Salyk}, C., {Pontoppidan}, K.~M., {Blake}, G.~A., {et~al.} 2008, \apjl, 676,
  L49

\bibitem[{{Sch{\"o}ier} {et~al.}(2005){Sch{\"o}ier}, {van der Tak}, {van
  Dishoeck}, \& {Black}}]{schoier2005}
{Sch{\"o}ier}, F.~L., {van der Tak}, F.~F.~S., {van Dishoeck}, E.~F., \&
  {Black}, J.~H. 2005, \aap, 432, 369

\bibitem[{{van Dishoeck} {et~al.}(2006){van Dishoeck}, {Jonkheid}, \& {van
  Hemert}}]{vd_diskphoto}
{van Dishoeck}, E.~F., {Jonkheid}, B., \& {van Hemert}, M.~C. 2006, in Faraday
  Discussions, Vol. 133, Faraday Discussions, 231

\bibitem[{{Weidenschilling} \& {Cuzzi}(1993)}]{wc_ppiii}
{Weidenschilling}, S.~J. \& {Cuzzi}, J.~N. 1993, in Protostars and Planets III,
  ed. E.~H. {Levy} \& J.~I. {Lunine}, 1031--1060

\bibitem[{{White} \& {Ghez}(2001)}]{wg01}
{White}, R.~J. \& {Ghez}, A.~M. 2001, \apj, 556, 265

\end{thebibliography}

\end{document}